\documentclass{elsart}
\usepackage{natbib,psfig}
\begin{document}
\def\arcsec{\hbox{$^{\prime\prime}$}}
\runauthor{Hoare M. G.}
\begin{frontmatter}
\title{Star Formation at High Angular Resolution}

\author{M. G. Hoare}

\address{School of Physics and Astronomy, University of
  Leeds, Leeds, UK}

\begin{abstract}
The role of the SKA in high resolution observations of the formation
of low and high mass stars in the Galaxy is examined.  The large 
collecting area will have a large impact on dynamical studies at
high resolution using spectral lines. The SKA will allow
dramatic progress in the investigation of the ionised,
atomic and molecular phases over a range of evolutionary stages.
Attention is focused here on the potential of radio recombination
lines, H~I 21 cm line and molecular Zeeman measurements. This
combination will unveil the physical processes that drive
jets and winds from young stars.  Other important areas such as
pre-biotic molecules, disc ablation in OB clusters and massive star
formation in nearby galaxies are also briefly discussed.
\end{abstract}

\end{frontmatter}

\section{Introduction}

The giant leap in sensitivity provided by the SKA and its large and
flexible range of capabilities will provide ways to answer
key problems in star formation in the Galaxy and in nearby galaxies.
The impact of the SKA on studies of the ionised, atomic and molecular
phases are considered in below.  Ionised gas occurs in jets,
stellar winds, shocked regions and H~II regions and is the most
obvious aspect where the SKA will bring vast new insight. The very high
continuum sensitivity of the SKA will enable new studies of weak
emission regions. However, in a galactic context the order of
magnitude increase in sensitivity that will become available with the
EVLA and e-MERLIN will have already delivered continuum maps of large
samples of jets and winds from young stars. These will cover a range
of masses, ages, inclination angles, etc., sufficient to permit firm
conclusions to be drawn vis-\`{a}-vis the geometry of the
mass-loss. The major unique capability that the SKA will bring is the
ability to trace the dynamics of these outflows via recombination
lines at high resolution.

The principal driver for the SKA is the ability to detect H I and this
will bear considerable fruit in star formation studies. This component
has hardly been explored at all so far due to lack of sensitivity at
the required resolution. A key question here is whether the high
spatial resolution will be able to cut through the enormous amount of
line-of-sight H I that be-devils galactic observations, especially
towards star forming regions. Most molecular studies will be provided by
ALMA and the SKA contribution here will be in terms of mapping the
magnetic fields that lie at the heart of the star formation process.

\section{Ionised Gas}

Outflows are likely to play a dominant role in angular momentum loss and
in setting the final mass of a star as the infalling material is
cleared away (e.g. Shu 2003). A wind interacts with the larger scale cloud
increasing the turbulence. The effects of the more massive stars
can both trigger further star formation (Elmegreen \& Lada 1977)  and
eventually disperse the cloud (Franco et
al. 1994). Observations of the kinematics of these outflows at
sufficient resolution is the key to understanding the physics
driving them. It is also important to catch these outflows at the
onset since this is where they will have most impact on the formation process
itself. This necessitates the study of the youngest and most
embedded objects and hence the radio regime is the only one that can
deliver extinction-free views of this process. There are good reasons
for believing that the physics behind the formation of low and high
mass stars is different.

\subsection{Low-mass star formation}

For low-mass stars the outflow is usually highly collimated in the
form of a jet (PP4 review). The general consensus is that these jets
are driven by magneto-hydrodynamic forces, but the actual controlling
processes are still far from clear. In particular, the geometry of the
magnetic field is unknown and even whether it is the field originating
in the star, accretion disc or the interaction of the two that is
important (Breitmoser \& Camenzind 2002, Ouyed \& Pudritz 1997, Shu \&
Shang 1997). The inner most regions of the disc are thought to be
disrupted by strong stellar fields with in-fall then occurring along
field lines. This has implications for the planetary formation
process. 

Since the magnetic field itself is difficult to observe in detail, the
best way to distinguish between these competing models is to observe
the velocity field in the outflow where it is launched and
collimated. Radio recombination lines provide the only means to do
this for the highly embedded young objects not seen directly even in
the mid- and near-IR - the so-called Class 0 objects (Andr\'{e} et
al. 1993). These are the ones with high in-fall and outflow rates
where much of the key star formation physics is occurring.  Currently
the VLA can detect and resolve the continuum emission of only the
bases of these dense jets (e.g. Figure 1a and Reipurth et al 1999).

\begin{figure}

\begin{center}\begin{minipage}{6cm}
\psfig{file=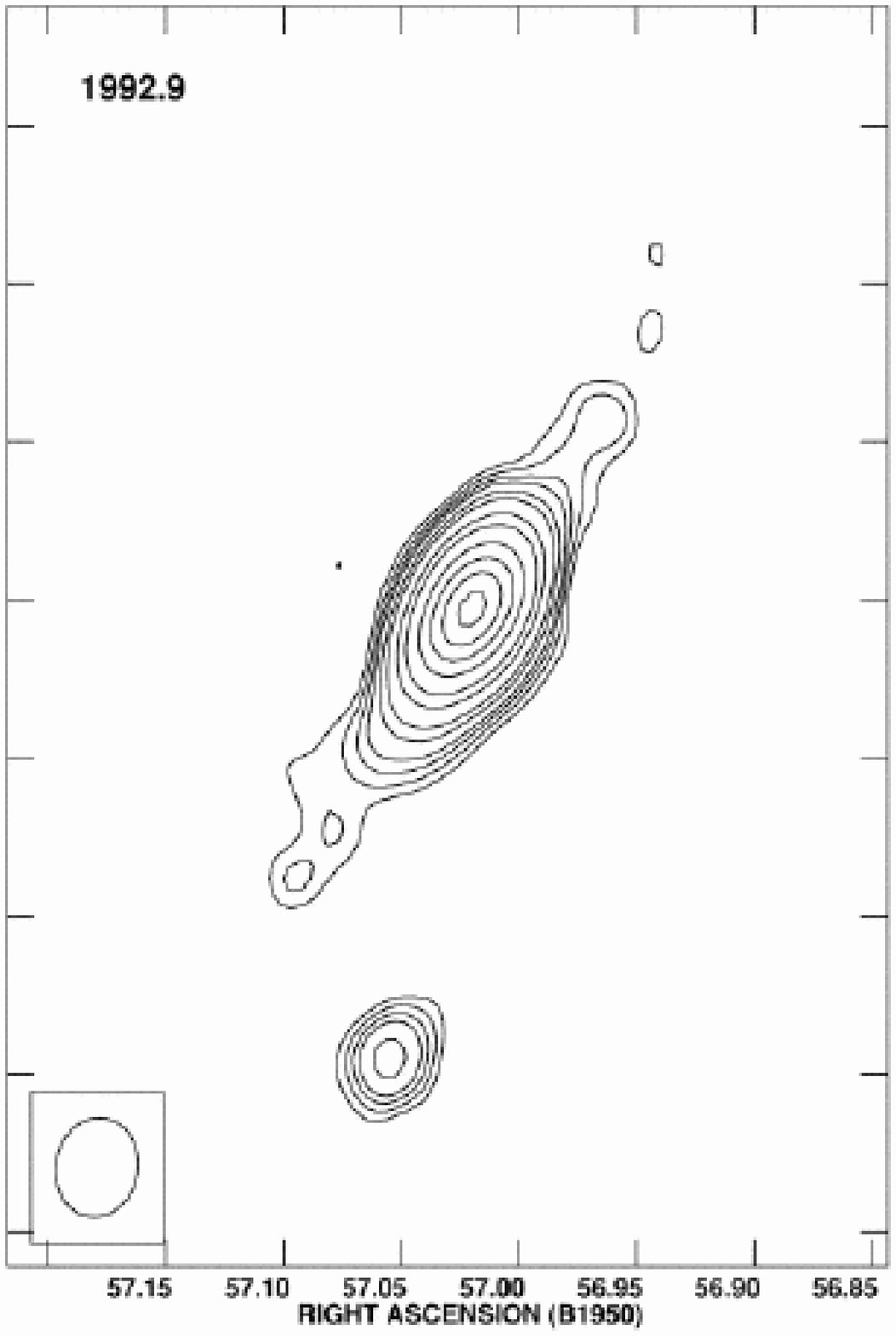,width=6cm}\end{minipage}\hspace*{1cm}\begin{minipage}{5.5cm}\psfig{file=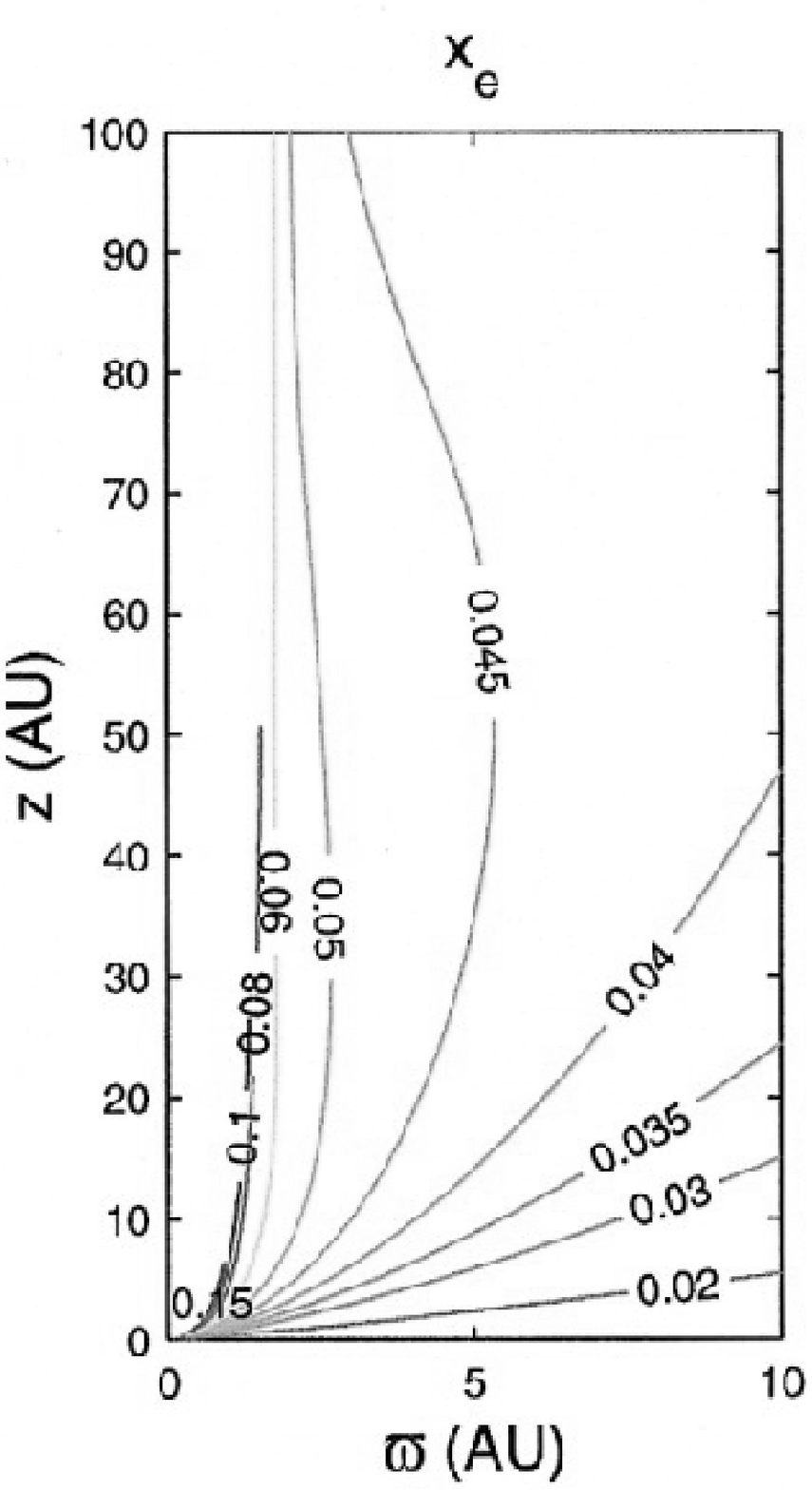,width=5.3cm}\end{minipage}
\end{center}

\caption{a) 6cm VLA continuum map of the radio jet VLA 1 that powers
  the HH 1 region at 0.5\arcsec\ resolution. The other unresolved
  source is VLA 2. The peak continuum brightness temperature is 100 K. From
  Rodr\'{i}guez et al. (2000). b) Predicted ionisation fraction
  structure for an X-wind model for YSO jets. Note the physical scale
  in AU assuming a distance of 150 pc appropriate for the nearest
  low-mass star-forming regions. From Shang et al. (2002).}
\end{figure}

Optical line observations are just beginning to resolve the velocity
structure in less embedded objects. Bacciotti et al. (2002) have used
STIS observations of the jet in DG Tau to reveal that a significant
amount of angular momentum is indeed removed by the jet.  The angular
resolution of the SKA of 7 mas at 20 GHz corresponds to 1 AU at the
distance of the nearest low-mass young stars in Taurus. This is just
the scale where the models predict significant acceleration and
collimation to occur.  At these scales the jet is usually obscured in
the optical and even the near-IR. Imaging interferometers operating in
these wavebands are the only other facility that will be able to probe
milli-arcsecond scales in the mean time. A comparison of detailed
studies of the velocity structure of the ionised jets from the SKA
with that seen in the molecular discs from ALMA observations will
allow a comprehensive theory of accretion, outflow and angular
momentum transport to be realised.

The more evolved T Tauri stars often exhibit variable non-thermal
radio emission which is sometimes associated with X-ray emission as
well (e.g. Feigelson et al 1998). This is thought to arise via
gyro-synchrotron emission linked to the magnetic flaring activity in
the star/disc magnetic field interaction. Strong magnetic events in
the solar nebula are needed to explain some features seen in
meteorites. The wide-field/multi-beaming capability of the SKA would
allow monitoring of a large number of targets to study the timescales
and energetics of this phenomenon. The high resolution of the SKA may
be able to localise the activity to the star or disc and test
predictions in this context of the X-wind model (Shu et al 2001). With
most star formation occurring in clusters there will be a large
multiplexing possible with many of the star formation studies
(e.g. Rodr\'{i}guez et al. 1999). The one degree field of view of the
SKA is easily sufficient for this.

A large proportion of low-mass stars form in clusters, which have
luminous, young OB stars at their centres (Carpenter et al. 2000). The UV radiation and
winds from the early-type stars are sufficiently strong to
photo-evaporate and destroy the circumstellar discs around the lower
mass stars. This could have a significant effect on the fraction of
stars that go on to form planetary systems. So far this ablation
process has only been seen in the nearest massive star-forming region
in Orion 0.5 kpc away (O'Dell \& Wen 1994). The ionised gas flowing
off the discs is seen in the radio continuum at the level
of a few mJy (Garay, Moran \& Reid 1987) which is spatially resolved
by MERLIN (Graham et al. 2002). This
disc destruction phase is likely to be short-lived and so a large
sample of evolved H~II regions, i.e. up to 10 times further away, will
need to be searched for this phenomenon to gauge its importance. This
is likely to require the sensitivity and resolution of the SKA.

\subsection{High-mass star formation}

Young high mass stars themselves also emit thermally at radio
wavelengths even before they begin to ionise the surrounding ISM
to form ultra-compact H~II regions.  A few of these have been resolved
to reveal highly collimated radio jets which are similar to their
low-mass counterparts (Mart\'{i} et al. 1998; Rodr\'{i}guez et
al. 1994). However, the very extended optical and near-IR jets seen in
low-mass star-forming regions are certainly not the norm (Poetzel et
al. 1992; Davis et al. 1998). Other high mass young
stellar objects have been shown to drive an equatorial outflow via
high resolution radio mapping (Figure 2 and Hoare 2002). This has been
interpreted as material being driven off the surface of an accretion
disc by the radiation pressure of the luminous young star (Drew et
al. 1998). When jets are seen in high-mass objects then it implies
that a magneto-hydrodynamic mechanism is at work as for low-mass
objects, although purely hydrodynamic collimation mechanisms have been
investigated (Mellema \& Frank 1997).  At present it is unclear
whether the dichotomy between jets and equatorial winds, and by
implication magnetic and radiative driving mechanisms, is a function
of evolutionary stage, mass, or initial conditions.

Radio continuum mapping of large samples with e-MERLIN and EVLA will
reveal which of these geometries is common and go some way to
answering these questions.  Recombination line mapping of the dynamics
with the SKA will test the disc ablation scenario and determine
whether this is important in setting the final mass and hence the
upper IMF. In the case of high mass jets the velocity structure should
distinguish between MHD and hydrodynamic collimation; the former
likely to impart more rotation into the jet.

\begin{figure}[t]

\centerline{\psfig{figure=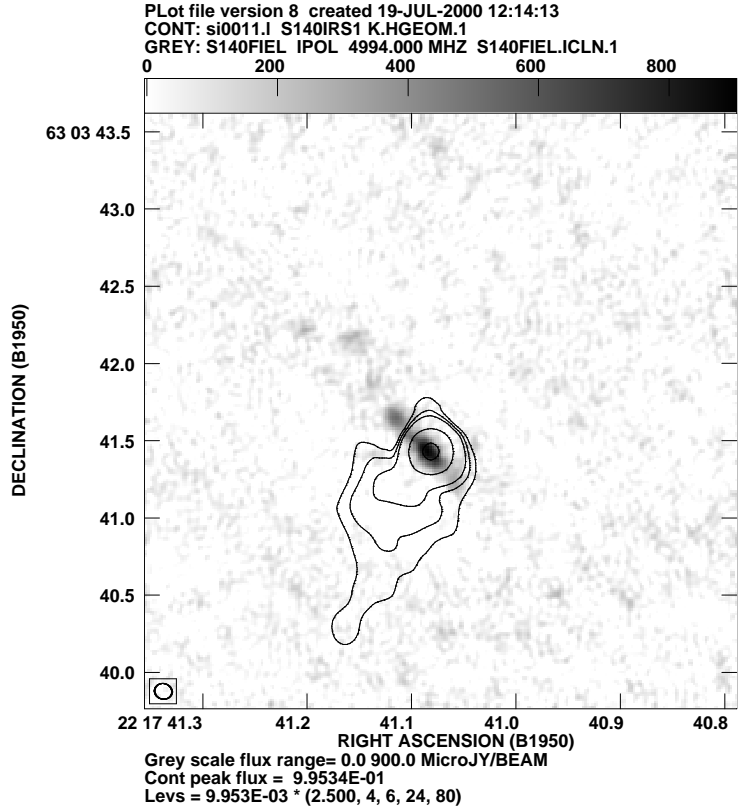,height=11cm}}

\caption{MERLIN 5 GHz image of the equatorial wind from the massive
young stellar object S140 IRS 1. Overlaid are contours of a high
resolution near-IR image showing the monopolar reflection nebula
arising from light scattered off the walls of the lobe of the bipolar
molecular outflow directed towards us. The SKA will be able to probe
the dynamics of these important mass-loss processes at resolutions at
least as high as this. From Hoare (2002).}
\end{figure}

The extreme continuum sensitivity of the SKA should allow the thermal
wind emission from massive young stellar objects to be detected in
near-by metal-poor galaxies like the Magallenic Clouds. Typical fluxes
of nearby ($\sim$1 kpc) examples are about 1 mJy, which corresponds to
300 nJy at 55 kpc. This will allow an investigation of the mass-loss
rate as a function of metallicity for a given luminosity. In
particular, if the line-driven equatorial wind scenario is confirmed
then the mass-loss will scale with metallicity as is being
demonstrated for main sequence OB stars (Crowther et al. 2002). Furthermore if
the ionised winds also play a role in setting the final mass of the
star this could then lead to the first concrete proof of a physical
mechanism whereby lower metallicity environments lead to more massive
stars.

A few very dense and compact H~II regions with rising spectral indices
have been found to have components in their radio recombination line
profiles broader than expected (Jaffe \& Mart\'{i}n-Pintado 1999;
Sewilo et al. 2004; De Pree et al. 2004). The large widths (60-100
km~s$^{-1}$) are not due to pressure broadening and must reflect the
dynamics. The widths are reminiscent of the even broader near-IR
recombination line profiles from the ionised winds from the massive
young stellar objects discussed above (e.g. Bunn et
al. 1995). However, their radio continuum brightnesses are much higher
than these pure ionised wind sources. 

Several of the very compact objects exhibit bi-polar morphologies like
one of the original sources in this class NGC 7538 IRS 1 (Gaume et
al. 1995). Photo-evaporating discs have therefore been suggested as a
possible explanation, but like compact H~II regions themselves these
only give rise to thermal gas motions and would therefore struggle to
explain such widths. Photo-evaporation also ignores the effect of
radiation pressure that explains the equatorial winds seen in some
massive YSOs. Perhaps these broad recombination objects represent a
transition phase from a massive YSO wind to ultra-compact H~II region.
A fast radiatively driven wind could mass-load sufficiently to produce
the bright radio emission at intermediate velocities. Gaume et
al. (1995) invoke this kind of picture for NGC 7538 IRS
1. Alternatively an extreme champagne-type flow could occur rather
like a small-scale version of the 'blow-out' proposed for one of the
larger scale broad recombination line objects S106 (Dyson 1983). As the
Lyman continuum radiation from the central star turns on ionised gas
would expand rapidly down the bipolar cavity previously evacuated by
the predominately molecular outflow.

It is the classical H~II regions that have traditionally been the
subject of radio investigations of massive star formation.  Most of
the remaining questions concerning galactic H~II regions such as their
dynamics and evolution are likely to have been solved before the SKA
begins operation, even for the ultra-compact ones (e.g. Lumsden \&
Hoare 1999). The new field opened up by the SKA will be to carry out
the kind of studies currently done on galactic H~II regions in nearby
spiral and irregular galaxies. Relatively face-on spirals such as M 33
present an excellent environment in which to examine the global
aspects of massive star formation as traced through their H~II region
populations. The SKA resolution of 17 mas at 5 GHz corresponds to
nearly 0.05 pc in M 33. This is the realm of ultra-compact H~II
regions, which are commonly still deeply embedded in their parent
molecular cloud and thus not accessible to optical or even near-IR
studies. Being young, they also have the most relevance to the
conditions in which the OB stars were born. Typical fluxes in M 33 for
such objects are tens of nJy even if powered by a B2V star. Hence, we
can probe the whole upper IMF. Currently, we are just skimming the top
of the IMF with the VLA in terms of the compact H~II region population
in M 33 (Figure 2 and Pattison \& Hoare, in prep.).

\begin{figure}[t]

\centerline{\psfig{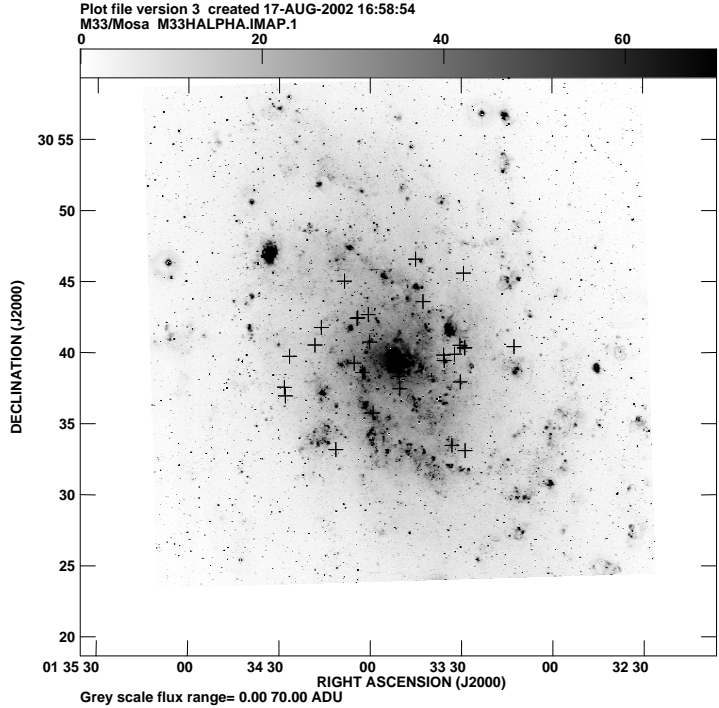}}

\caption{Crosses mark the locations of young, dense H~II regions found
in a 2 hour 5 GHz VLA A configuration observation overlaid on an
H$\alpha$ image of the nearest relatively face-on spiral M 33. From
Pattison \& Hoare, in prep.}
\end{figure}

For the brighter objects it should be possible to again use the
collecting area of the SKA to detect their recombination lines. This
is doable now for galactic H~II regions at high resolution using the
VLA (e.g. Wood \& Churchwell 1991). The power of such velocity studies
with the SKA will be in combination with CO studies from ALMA and H~I
with the SKA to build up a 3D picture of where massive stars are
forming. This can then be related to possible triggers of
gravitational collapse such as the spiral density wave, clusters of
more evolved OB stars and supernova remnants. Such a dataset can
discriminate between density wave and self-propagating models of
spiral structure. The question of what triggers massive star formation
is much easier to answer in nearby spirals than in our own galaxy due
to the problems of everything being along the same line-of-sight in
the galactic plane.

\subsection{Recombination line mapping}

The big step forward from studies of the ionised gas in galactic star
formation with the SKA will come from probing dynamics at the highest
resolution.  For stellar wind sources a crude estimate of the strength
of the recombination lines can be made from the ratio of the line
brightness temperatures from the optically thin part of the wind to
the continuum brightness temperature. It has been observed that even
when a source is optically thick in the continuum, recombination lines
at the same frequency are still easily detectable from the surrounding
diffuse envelope (Sewilo et al 2004). Altenhoff et al. (1981) show for
a constant velocity wind that typical line-to-continuum brightness
temperature ratios are of order 0.1.  The continuum brightness
temperatures seen at current resolutions in stellar jets and winds are
in the range of 10$^{2-4}$ K (Rodriguez 1999; Hoare et al. 1994; Hoare
\& Muxlow 1996). At higher resolution we can expect higher continuum
brightness temperatures in the regions of interest and 
so the line brightness temperatures will be about 100 K.

This is currently the sensitivity of the VLA in a 10 km s$^{-1}$
channel after a 12 hr integration with 0.1\arcsec\ resolution at 20
GHz. Gaume et al. (1995) used this setup to resolve the very strong,
broad recombination lines from the young high-mass source NGC 7538 IRS
1. Hence, the SKA, with two orders of magnitude better line
sensitivity, can map these sources at 10 times better resolution and
hence probe the dynamics at interesting scales.

Many aspects of recombination line physics (see Gordon \& Sorochenko
2002) drive the SKA requirements to the highest frequencies apart from
the need for the highest spatial resolution. Firstly
the lines themselves get stronger with increasing frequency. 
The line-to-continuum ratio of the H66$\alpha$ line at 22.4 GHz
is at least an order of magnitude higher than that for H166$\alpha$
at 1.4 GHz. Another severe problem for lower frequency lines is that of
pressure broadening. The ratio of pressure broadening $\Delta v^{l}$
to Doppler broadening $\Delta v^{D}$ is given by
\begin{equation}
\frac{\Delta v^{l}}{\Delta
  v^{D}}=0.14\left( \frac{n}{100} \right)^{7.4}\left(
  \frac{n_{e}}{10^{4}} \right)
\end{equation}
where $n$ is the principal quantum number and $n_{e}$ is the electron
density (Keto et al 1995; De Pree et al 2004).  Hence, for the highest
frequency line likely to be observed by the SKA (H66$\alpha$) only
densities up to about 10$^{6}$ cm$^{-3}$ can be observed before
pressure broadening begins to dominate. When it does dominate the
lines will become much more difficult to detect and less useful as
probes of the dynamics.  Martin (1996) deduced that the electron
density at 1 AU from the star for typical T Tauri star jets is at
least 10$^{7}$ cm$^{-3}$ and could be much higher. However, in any
kind of wind the density falls off very rapidly, usually as $r^{-2}$,
and so parts of the wind will still produce narrow components to the
line profile that are good tracers of the kinematics.  The sensitivity
to density via the pressure broadening at the transition between
thermal and pressure broadening, of course also allows an accurate
measure of density which will be another valuable probe.

A possible disadvantage of higher frequency recombination lines is
their susceptibility to stimulated emission and maser emission. This
is seen in the highest frequency recombination lines in the millimetre
regime in the wind of the enigmatic object MWC349A, whose nature is
not fully understood (Mart\'{i}n-Pintado et al 1993). Again this would
make interpretation of the velocity and density structure more
difficult. Non-LTE effects are also common in radio recombination
lines. Mart\'{i}n-Pintado et al (1993) and Jaffe \& Mart\'{i}n-Pintado
(1999) attribute the broad flat-topped cm lines to non-LTE effects.

With the large bandwidth and a flexible correlator it will also be
possible to observe about 10 recombination lines simultaneously
enabling constraints on the velocity and density structure through one
observations as well as increasing the significance of line detections
overall. Multiple lines also help disentangle the effects of
stimulated emission and departures from LTE. Detailed modelling of the
strengths and profiles of radio recombination lines needs to be done
for stellar jets and winds to fully appreciate what the SKA will and
will not be able to do in this field.

\section{Atomic Gas}

The very high sensitivity of the SKA to atomic hydrogen via the 21 cm
line will also provide unique new insights into the star formation
process. Again this is most likely to come in the realm of outflow
studies, not because there is not much atomic gas involved, but rather
because there is too much. The star forming clouds will have
significant atomic layers both near the young stars as they dissociate
the dominant molecular component and around the edges of the
complex. This together with the abundance of H~I along the line of
sight to most star forming regions in the galaxy presents a vast
confusion problem at velocities close to the systemic velocities of
forming stars. The high spatial resolution of the SKA will certainly
resolve out much of the confusing gas, but it is most likely to be the
spatial resolution combined with high velocity signatures that will
give clear new insights into the physics of star formation. As is
clear from Figure 1 only a small fraction of the mass-loss in stellar
jets from low-mass young stars is ionised. Most of the mass of these
jets is therefore likely to be in atomic form, although some may also
be molecular. Shocked molecular hydrogen emission associated with 
the jet 
has been seen in a few cases (e.g. McCaughrean et al 1994).

Atomic hydrogen that can be directly associated with outflows from
young stars has only been seen in a couple of cases so far. Lizano et
al. (1988) H~I emission wings up to 170 km~s$^{-1}$ in towards the
exciting source of HH 7-11 using Arecibo. The derived mass-loss rate
of 3$\times$10$^{-6}$ M$_{\odot}$yr$^{-1}$ is sufficient to drive the
CO flow from this object. Rodr\'{i}guez et al (1990) used the VLA to
resolve the intermediate velocity H~I into a bipolar flow similar to
the known CO flow.  Extended intermediate velocity bipolar H~I
emission has also been detected at about arcminute resolution from
L1551 by Giovanardi et al (2000). Higher velocities still were seen in
Arecibo observations of L1551 by Giovanardi et al (1992).  Giovanardi
et al (2000) modelled the L1551 emission as a bi-conical interaction
zone between a high speed wind and walls of the bipolar molecular
flow.

Current facilities are unable to pick up any H~I emission directly
from the highly collimated jets, but this will be a key aim of SKA
studies. The resolution of 100 mas at 1.4 GHz is unlikely to be
sufficient to probe right into the acceleration and collimation zones
as the high frequency recombination line studies will. However, H~I
will reveal the velocity structure in the bulk of the mass of the flow
and so should yield further important constraints on the mass-loss
mechanism.  If the density in the jet is about 10$^{6}$ cm$^{-3}$ and
it is 100 AU in diameter then the H~I column density is around 10$^{21}$
cm$^{-2}$.  For emission spread across 10 km~s$^{-1}$ this 
corresponds to an antenna temperature of 100 K which is within the
capabilities of the SKA operating at its highest angular resolution.
These kind of jet parameters also give significant H I optical depth
against the bright continuum emission from the ionised portion of the
jet. The well-defined geometry of absorption will yield additional
constraints on jet models.

The SKA H~I resolution of 15 AU at Taurus would be able to resolve the
surface atomic layer of the accretion discs themselves if the ambient
and line-of-sight material can be sufficiently resolved out. This will
allow the dynamics of this interface zone between the Keplerian outer
disc (studied with ALMA) and the radiation and flows from the inner
disc region to be studied. It is here where the clearing of the
infalling envelope and launching of any wind-angled wind may take
place.

If suitable narrow velocity components can be found in the dense
ambient atomic gas then Zeeman measurements may be possible. These
could be in emission or in absorption against the weak continuum,
analogous to what is done against strong H II region continua at
present (Crutcher 1999). This would enable the magnetic field strength
and line-of-sight geometry to studied at high resolution. Continuous
spatial measurements would have much greater diagnostic power than the
sparse and possibly special locations currently made using OH masers.

\section{Molecular Gas}

By the time the SKA begins operations ALMA will have revolutionised
our view of the molecular gas at high resolution in star forming
regions. The detailed molecular observations will be the setting in
which the ionised and atomic dynamical studies by the SKA described
above will take place. However, the SKA will also make important
contributions to our understanding of the molecular environment
through high resolution magnetic field measurements. With nearly
every aspect of star formation, certainly for low-mass stars, thought
to be controlled by magnetic processes the importance of this cannot
be understated.

Although progress will have been made in the use of millimetre-wave
transitions of molecules such as CN for Zeeman splitting measurements
using ALMA (Crutcher et al. 1999) there are advantages to using
cm-band transitions.  The large Zeeman splitting of molecules such as
CCS holds great promise for detecting weak fields or mapping strong
ones at high resolution (Levin et al 2001). The cm transitions of
heavy molecules also have the advantage of reduced thermal line widths
being able to penetrate deep into dense cores without the dust
becoming optically thick.  The combination of measurements from
emission and absorption against the continuum from these and the
traditional Zeeman transitions of H I and OH will help reveal the
geometry of the magnetic field.  The velocity information from either
the thermal Zeeman lines or maser measurements, which will be much
more numerous than at present, will enable a three-dimensional picture
of the magnetic field strength to be built up. Polarisation maps from
ALMA will also help pin down the geometry. This information is another
vital part of testing the models for mass-loss and angular momentum
transport in young disc/outflow systems.

The extreme sensitivity of the SKA will also allow a veritable
pin-cushion of extragalactic sources to be seen through dense cores
before and after the onset of collapse. Absorption magnetic field
measurements across the cores will be key to understanding the role of
magnetic support and ambipolar diffusion in these early stages of star
formation. 

One final area where the SKA will make a great impact is in the
detection of heavy molecules and in particular pre-biotic
molecules. As ever more complex molecules form in the densest parts of
the proto-planetary discs their main transitions inevitably move from
mm to cm wavelengths.  Models of proto-planetary discs are now being
developed which provide a two-dimensional axisymmetric solution of the
coupled physical and astro-chemical problem to predict molecular
distributions in the inner 10 AU of the disk (Ilgner et al. 2004).
The upper layers of the discs where much of the line formation takes
place require a full treatment of the ionisation and dissociation
fronts. Chemical models of complex organic molecule synthesis will
then predict the concentrations and distributions of biologically
relevant species such as glycine, adenine and other DNA bases. Many of
these species have low-lying rotational transitions at frequencies of
less than 20 GHz.  Hence, they SKA will provide the ideal capability
to map out the development of pre-biotic molecules during the later
stages of the star formation process.

\section{Conclusions}

The SKA will make valuable contributions to our understanding of star
formation through its ability to probe the dynamics of ionised and
atomic gas at high resolution. Coupled with the SKA's pre-eminence in
magnetic field mapping it will enable the driving mechanism of
collimated and equatorial mass-loss from low and high mass young
stellar objects to be firmly established. Key pieces in the planet
formation puzzle will be put in place through studies of the
disruption of the disc by stellar magnetic fields and
photo-evaporation by near-by OB stars.  Galactic H~II region studies
will be transfered into nearby galaxies where the combination of SKA
and ALMA will allow the global and local triggering of high-mass star
formation to be established. The possibility of directly
measuring the metallicity dependence of mass-loss from young stars
will have significant implications for our understanding of star
formation at high redshift. 

\section{Acknowledgements}

I am indebted to Lewis Knee and Tom Millar for input and Sean
Dougherty, Steve Rawlings and Chris Carilli for their encouragement and
infinite patience.

\end{document}